\documentclass[aps,prl,reprint,letterpaper,superscriptaddress,showpacs]{revtex4-1}
\usepackage{amssymb}
\usepackage{keyval}%
\usepackage{graphicx}
\usepackage{dcolumn}
\usepackage{bm}
\usepackage{color}
\usepackage{fancyhdr}

\newcommand{\be}{\begin{equation}}
\newcommand{\ee}{\end{equation}}
\newcommand{\bea}{\begin{eqnarray}}
\newcommand{\eea}{\end{eqnarray}}

\newcommand{\ignore}[1]{}

\ignore{
\pagestyle{fancy} %
\pagestyle{fancyplain} %
\lhead{Yin, Wei-Guo} %
\chead{\sl DRAFT --- NOT FOR CIRCULATION\vspace{-2pt}}
\rhead{\today} %
\cfoot{\sc\thepage} %
\lfoot{} %
\rfoot{} %
}

\begin{document}
\title{
Ferromagnetic Exchange Anisotropy from Antiferromagnetic Superexchange\\ in the Mixed $3d-5d$ Transition-Metal Compound  Sr$_3$CuIrO$_6$}

\author{Wei-Guo Yin}
\email{wyin@bnl.gov}
\affiliation{Condensed Matter Physics and Materials Science Department, Brookhaven National Laboratory, Upton, New York 11973, USA\\}
\author{X. Liu}
\affiliation{Beijing National Laboratory for Condensed Matter Physics, and Institute of Physics, Chinese Academy of Sciences, Beijing 100190, China\\}
\affiliation{Condensed Matter Physics and Materials Science Department, Brookhaven National Laboratory, Upton, New York 11973, USA\\}
\author{A. M. Tsvelik}
\affiliation{Condensed Matter Physics and Materials Science Department, Brookhaven National Laboratory, Upton, New York 11973, USA\\}
\author{M. P. M. Dean}
\affiliation{Condensed Matter Physics and Materials Science Department, Brookhaven National Laboratory, Upton, New York 11973, USA\\}
\author{M. H. Upton}
\affiliation{Advanced Photon Source, Argonne National Laboratory, Argonne, Illinois 60439, USA\\}
\author{Jungho Kim}
\affiliation{Advanced Photon Source, Argonne National Laboratory, Argonne, Illinois 60439, USA\\}
\author{D. Casa}
\affiliation{Advanced Photon Source, Argonne National Laboratory, Argonne, Illinois 60439, USA\\}
\author{A. Said}
\affiliation{Advanced Photon Source, Argonne National Laboratory, Argonne, Illinois 60439, USA\\}
\author{T. Gog}
\affiliation{Advanced Photon Source, Argonne National Laboratory, Argonne, Illinois 60439, USA\\}
\author{T. F. Qi}
\affiliation{Center for Advanced Materials, University of Kentucky, Lexington, Kentucky 40506, USA\\}
\affiliation{Department of Physics and Astronomy, University of Kentucky, Lexington, Kentucky 40506, USA\\}
\author{G. Cao}
\affiliation{Center for Advanced Materials, University of Kentucky, Lexington, Kentucky 40506, USA\\}
\affiliation{Department of Physics and Astronomy, University of Kentucky, Lexington, Kentucky 40506, USA\\}
\author{J. P. Hill}
\affiliation{Condensed Matter Physics and Materials Science Department, Brookhaven National Laboratory, Upton, New York 11973, USA\\}

\date{\today}

\begin{abstract}
We report a combined experimental and theoretical study of the unusual ferromagnetism in the one-dimensional copper-iridium oxide Sr$_3$CuIrO$_6$. Utilizing Ir $L_3$ edge resonant inelastic x-ray scattering, we reveal a large gap magnetic excitation spectrum. We find that it is caused by an unusual exchange anisotropy generating mechanism, namely, strong ferromagnetic anisotropy arising from antiferromagnetic superexchange, driven by the alternating strong and weak spin-orbit coupling on the $5d$ Ir and $3d$ Cu magnetic ions, respectively. From symmetry consideration, this novel mechanism is generally present in systems with edge-sharing Cu$^{2+}$O$_4$ plaquettes and Ir$^{4+}$O$_6$ octahedra. Our results point to unusual magnetic behavior to be expected in mixed $3d-5d$ transition-metal compounds via exchange pathways that are absent in pure $3d$ or $5d$ compounds.
\end{abstract}

\maketitle
\sloppy

The interest in strongly correlated electronic systems has recently been extended from $3d$ transition-metal compounds (TMCs) to $5d$ compounds. Usually, the strength of electron correlation is characterized by the ratio of the local Coulomb repulsion to the electronic bandwidth. A large value of this ratio ($\sim 8$) is common in $3d$ TMCs such as superconducting cuprates \cite{review:Dagotto_science05}. Since $5d$ orbitals are more extended in space and host weaker Coulomb interactions than $3d$ orbitals, $5d$ TMCs are generally expected to be weakly correlated. However, it has been pointed out \cite{Sr2IrO4:Kim_Science09,Sr2IrO4:Kim_08,Sr2IrO4:Moon} that the relative weakness of the Coulomb interaction is offset by the strong spin-orbit coupling (SOC), which is typically $\sim 0.5$ eV for $5d$ elements. This strong SOC leads to a significant splitting and narrowing of the electronic bands, and pushes $5d$ TMCs toward the strongly correlated regime. Indeed, the SOC-driven Mott metal-insulator transition was shown to exist in a variety of $5d$ iridium oxides \cite{Sr2IrO4:Kim_Science09,Sr2IrO4:Kim_08,Sr2IrO4:Moon,Sr2IrO4:magnon:Kim,%
A2Ir2O7:Pesin,Ir:Jackeli,Na2IrO3:Singh,Na2IrO3:zigzag:Liu,Na2IrO3:zigzag:Choi,Na2IrO3:zigzag:Ye,Sr2IrO4:spin-flop:Kim,
Sr2IrO4:Arita,Sr3CuIrO6:d-d:Liu}.
An important consequence of this is the entanglement of the orbital and spin degrees of freedom in the resulting localized magnetic moments (termed ``isospins''), which can lead to unusual superexchange pathways and to new physics, for example, the proposed spin-liquid state as encoded in the Kitaev model \cite{Kitaev:Kitaev} in the honeycomb-lattice (Li,Na)$_2$IrO$_3$ \cite{Ir:Jackeli} and possible superconductivity in the  square-lattice Sr$_2$IrO$_4$, which shows similar magnetic ordering and dynamics to the cuprates \cite{Sr2IrO4:magnon:Kim}.

\begin{figure}[t]
\includegraphics[width=\columnwidth,clip=true,angle=0]{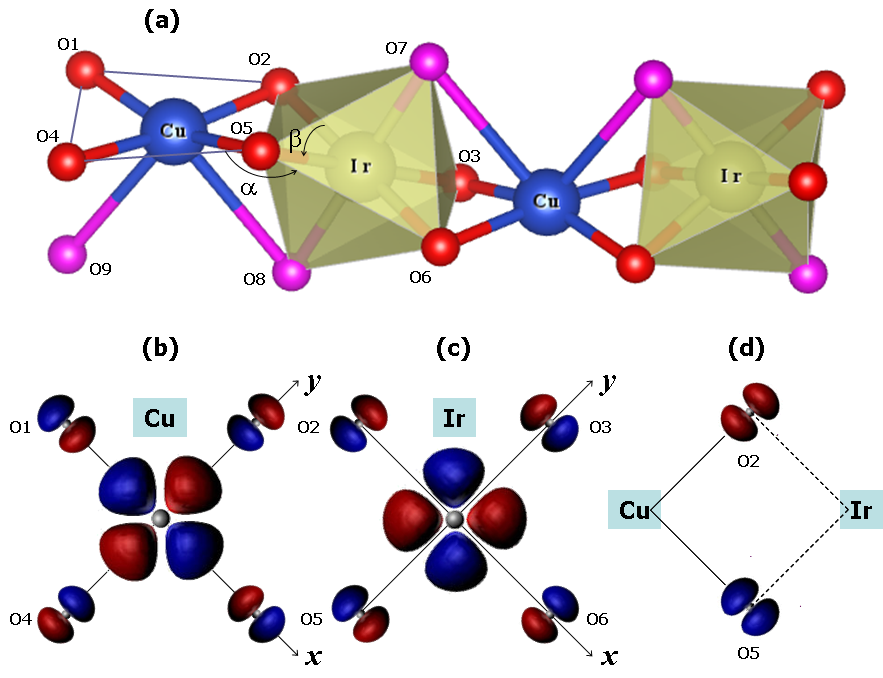}
\caption{\label{fig:orbital}(a) Cu-Ir chain of Sr$_3$CuIrO$_6$ where Cu$^{2+}$ and Ir$^{4+}$ are coordinated by an oxygen plaquette and octahedron, respectively. The IrO$_6$ octahedral tilting is denoted by $\alpha\simeq 150^\circ$ and the octahedral distortion by $\beta\simeq 82^\circ$. In the ideal case of $\alpha=180^\circ$ and $\beta=90^\circ$, the O2 and O5 and their neighboring Cu and Ir atoms form a square. (b),(c) Schematic drawings of Cu $3d_{x^2-y^2}$ and Ir $5d_{xy}$ Wannier orbitals $\phi_\mathrm{Cu}$ and $\phi_{\mathrm{Ir},xy}$, respectively. Note the considerable tails on the oxygen sites due to the metal-oxygen hybridization. (d) Schematic map of the overlap density $\rho=\phi_\mathrm{Cu}\phi_{\mathrm{Ir},xy}$. Red (blue) represents positive (negative) values.}
\end{figure}

The purpose of this Letter is to demonstrate that materials containing \emph{both} $3d$ and $5d$ magnetic ions can host new physics absent in either pure $3d$ or pure $5d$ compounds---because of the unique combination of unusual exchange pathways and special geometries, a result of the strong SOC of the $5d$ electrons and the differing coordinations of the $3d$ and $5d$ sites. Such materials will offer new avenues to engineer exotic magnetic behavior. To demonstrate this, we have chosen to study the one-dimensional copper-iridium oxide Sr$_3$CuIrO$_6$, as a prototype for such mixed $3d-5d$ systems. The  crystal structure of this compound [Fig.~\ref{fig:orbital}(a)] is reminiscent of both the superconducting cuprates and the iridium-based Mott insulators. Specifically, it contains chains of alternating Cu and Ir with the Cu$^{2+}$ in a planar oxygen coordination and the Ir$^{4+}$ in an octahedral oxygen coordination. The physical novelty of this material is demonstrated by the emergence of  ferromagnetic order, which is rare in pure cuprates or iridates (the closely related materials Sr$_3$CuPtO$_6$ and Sr$_3$ZnIrO$_6$ are antiferromagnetic) and was used to synthesize random quantum spin chain paramagnetism in Sr$_3$CuIr$_{1-x}$Pt$_x$O$_6$ \cite{Sr3Cu(IrPt)O6:Nguyen,Sr3(NiCuZn)IrO6:spin-pierls:Nguyen}. There is to date no microscopic understanding of this unique phenomenon of ferromagnetism.

Here, we use Ir $L_3$ edge resonant inelastic x-ray scattering (RIXS) \cite{RIXS:review:Ament,Sr2IrO4:magnon:Kim,Sr3CuIrO6:d-d:Liu,RIXS:La2CuO4:Dean} to reveal a large gap magnetic excitation spectrum in Sr$_3$CuIrO$_6$ and show that it is well described by an effective $S=1/2$ ferromagnetic Heisenberg model with an Ising-like exchange anisotropy.  We present a microscopic derivation of this model and find that the arrangement of alternating isospins and real spins on the edge-sharing Ir$^{4+}$O$_6$-Cu$^{2+}$O$_4$ chain leads to an unexpected effect, namely, that the ferromagnetic anisotropy arises from the antiferromagnetic superexchange. Our results point to an entirely new class of magnetic behavior in mixed $3d-5d$ TMCs.

\textit{RIXS measurements.}---The energy and momentum dependence of the magnetic excitation spectrum in a self-flux-grown Sr$_3$CuIrO$_6$ single crystal was studied by using Ir $L_3$ edge RIXS, for which dipole transitions excite and deexcite a $2p_{3/2}$ core electron to the $5d$ orbitals \cite{Sr2IrO4:magnon:Kim,Sr3CuIrO6:d-d:Liu,RIXS:review:Ament}. The measurements were carried out at beamline 9-ID, Advanced Photon Source, in a horizontal scattering geometry. A Si(844) secondary monochromator and a $R=2$ m Si(844) diced analyzer were utilized. The overall energy resolution of this setup was $\sim 45$ meV (FWHM) and the momentum resolution was better than 0.07 of the Brillouin zone (BZ) length along the Cu-Ir chain direction. All data were collected at 7 K.

\begin{figure}[t]
\includegraphics[width=\columnwidth,clip=true,angle=0]{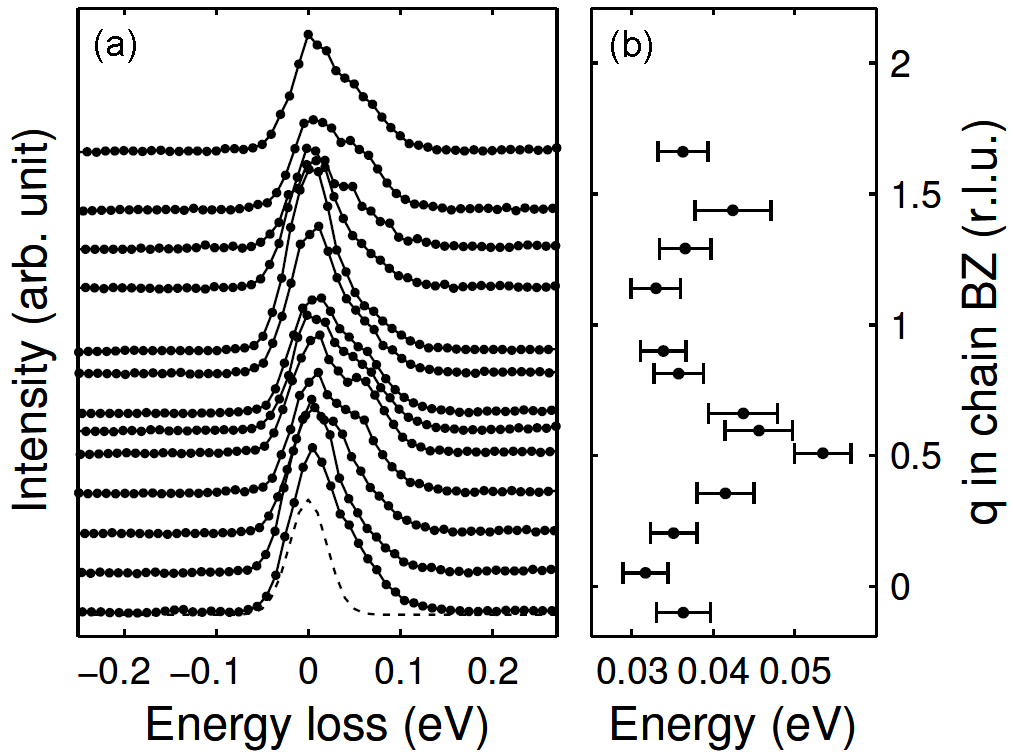}
\caption{\label{fig:RIXS}(a) RIXS spectra for different momentum transfer along the chain direction, compared with the off-resonant elastic line (dotted line).
(b) The extracted magnetic dispersion from the scans
shown in (a). The scans in (a) are offset to match
with their wave vector transfer along the chain direction presented in
(b).}
\end{figure}

Figure~\ref{fig:RIXS}(a) shows the low-energy RIXS spectra for the momentum transfer $q$ along the chain direction in the BZ in which the unit cell contains one Cu ion and one Ir ion. A shoulder appears near the elastic line and disperses along the chain direction. No dispersion is seen in this feature for momentum transfers perpendicular to the chain direction (not shown). We attribute this shoulder to a one-dimensional magnon mode. Its dispersion is plotted in Fig.~\ref{fig:RIXS}(b) and shows two salient features: (i) The minimum energy located at the $\Gamma$ point is not zero but rather shows a large gap of $\sim 30$ meV, comparable to the magnon bandwidth; (ii) the dispersive behavior follows the periodicity of the (one Ir plus one Cu) chain unit cell very well, approximately in the form of $\cos (qa)$, where $a$ is the nearest Ir-Ir distance.

These observations cannot be explained by conventional models. The simple linear chain ferromagnetic Heisenberg model proposed for this system \cite{Sr3Cu(IrPt)O6:Nguyen} is gapless.
In the related iridate Sr$_3$ZnIrO$_6$, two mechanisms have been proposed for a possible gap there. However, neither can be at work here. In the first, singlet dimers were proposed based on the alternating chain antiferromagnetic Heisenberg model \cite{Sr3(NiCuZn)IrO6:spin-pierls:Nguyen}. However, this cannot occur in Sr$_3$CuIrO$_6$, since it is ferromagnetic. In the second, an Ising-like exchange anisotropy is added to the
$S=1/2$ Heisenberg model \cite{Sr3ZnIrO6:XXZ:Lampe-Onnerud}. However, the magnon dispersion of the ferromagnetic model of this type, in which one does not distinguish between Cu and Ir atoms (i.e., the unit cell contains one magnetic ion), is in the form of $\cos (qa/2)$ \cite{spin:Yang,spin:Cloizer}, in disagreement with the dispersion observed above. Our data thus
require unconventional magnetism in this system and make a general call for a new understanding of systems with mixed $3d-5d$ magnetic ions.

\textit{Microscopic mechanism.}---To understand the microscopic origin of the ferromagnetism and the anisotropy in this system, we start by analyzing the geometry and symmetry of the Cu-Ir chain.

A portion of a Cu-Ir chain of Sr$_3$CuIrO$_6$ is shown in Fig.~\ref{fig:orbital}(a). The Cu$^{2+}$ ion is located at the center of an oxygen plaquette. The only magnetic orbital $\phi_\mathrm{Cu}$ centered on a Cu$^{2+}$ ion is of $x^2-y^2$ symmetry and is antisymmetric with regard to the Cu-Ir mirror plane [Fig.~\ref{fig:orbital}(b)].

The Ir$^{4+}$ ion lies in the center of an oxygen octahedron which shares an edge with neighboring CuO$_4$ plaquettes and experiences tilting and an orthorhombic distortion characterized by angles $\alpha\simeq 150^\circ$ and $\beta\simeq 82^\circ$, respectively, as defined in Fig.~\ref{fig:orbital}(a).
The octahedral ligand field renders the three Ir $t_{2g}$ ($5d_{xy}$, $5d_{xz}$, and $5d_{yz}$) orbitals relevant to the low-energy physics, while the two unoccupied Ir $e_g$ energy levels are $\sim 3$ eV higher \cite{Sr3CuIrO6:d-d:Liu}. We shall show below that the octahedral tilting and distortion are also relevant to the magnetism in Sr$_3$CuIrO$_6$ and that the role of the Ir $5d_{xy}$ orbital is completely different from those of the Ir $5d_{xz}$ and $5d_{yz}$ orbitals.

The Ir $5d_{xy}$ orbital $\phi_{\mathrm{Ir},xy}$ is symmetric with regard to the Cu-Ir mirror plane [Fig.~\ref{fig:orbital}(c)] and thus orthogonal to $\phi_\mathrm{Cu}$, even in the presence of the octahedral tilting and distortion. As a result, electron hopping between these two orbitals is prohibited and so is the superexchange process. Thus, the leading magnetic interaction between them is the direct exchange interaction $J_\mathrm{F}$, which is ferromagnetic \cite{spin:Kanamori,spin:Anderson}. From the measured magnon bandwidth (Fig.~\ref{fig:RIXS}), we conclude that $J_\mathrm{F}$ is of the order of dozens of meV. This is surprising, since direct exchange in TMCs is usually very small. The unusually large $J_\mathrm{F}$ comes from the fact that the tails of $\phi_{\mathrm{Ir},xy}$ and $\phi_\mathrm{Cu}$ actually overlap well around each of O2 and O5 in Fig.~\ref{fig:orbital}(d) \cite{Sr3CuIrO6:supplement}, similar to the large direct exchange ferromagnetism proposed for a Cu$^{2+}$-(VO)$^{2+}$ heterobinuclear complex \cite{Cu-VO:Kahn}.

By contrast, the Ir $5d_{xz}$ ($5d_{yz}$) orbital has a tail of $p_z$ symmetry around O2 (O5); its overlap with  $\phi_\mathrm{Cu}$ is vanishing everywhere, leading to a negligible direct exchange. Actually, the Ir $5d_{xz}$/$5d_{yz}$ orbitals are not orthogonal to $\phi_\mathrm{Cu}$ when $\alpha$ is tilted away from $180^\circ$, and electron hopping between them is allowed. 
Therefore, the leading magnetic interaction between them is the antiferromagnetic superexchange.

The above considerations lead to the following effective magnetic Hamiltonian for Sr$_3$CuIrO$_6$:
\begin{eqnarray}
 H = &\;& -J_\mathrm{F} \sum_{\langle \mathbf{m},\mathbf{n}\rangle} \vec{S}_{\mathbf{m},x^2-y^2} \cdot \vec{S}_{\mathbf{n},xy} \nonumber \\
 &+& J_\mathrm{AF} \sum_{\langle \mathbf{m},\mathbf{n}\rangle} \vec{S}_{\mathbf{m},x^2-y^2} \cdot (\vec{S}_{\mathbf{n},yz}+\vec{S}_{\mathbf{n},zx}) \nonumber \\
 & + & \lambda \sum_{\mathbf{n}} \vec{L}_\mathbf{n}\cdot \vec{S}_\mathbf{n}
  + \Delta \sum_{\mathbf{n}\sigma} d_{\mathbf{n},xy,\sigma}^\dag d_{\mathbf{n},xy,\sigma}^{}, \label{model:t2g}
\end{eqnarray}
where $\mathbf{m}$ denotes a Cu site, $\mathbf{n}$ an Ir site, and $\langle \mathbf{m},\mathbf{n}\rangle$ means nearest neighbors. There is one hole at each site, as implied by Cu$^{2+}$ and Ir$^{4+}$. $\vec{S}_{\mathbf{n},\gamma}=\sum_{\mu\nu}{d_{\mathbf{n},\gamma,\mu}^\dag \vec{\sigma}_{\mu\nu}^{} d_{\mathbf{n},\gamma,\nu}^{}}/2$ where $\vec{\sigma}_{\mu\nu}$ is the Pauli matrix and $d_{\mathbf{n},\gamma,\mu}$ is the annihilation operator of an electron with spin $\mu=\uparrow, \downarrow$ (or $\pm$) and orbital $\gamma=xy,xz,yz$ on the Ir site $\mathbf{n}$. $\lambda$ is the SOC constant on the Ir sites and $\Delta$ is the strength of the IrO$_6$ octahedral distortion.
$-J_\mathrm{F}<0$ is the ferromagnetic direct exchange coupling between the Ir $5d_{xy}$ and Cu $3d_{x^2-y^2}$ orbitals. $J_\mathrm{AF}>0$ is the antiferromagnetic superexchange coupling between the Ir $5d_{xz}$/$5d_{yz}$ and Cu $3d_{x^2-y^2}$ orbitals.

\begin{figure}[b]
\includegraphics[width=0.6\columnwidth,clip=true,angle=0]{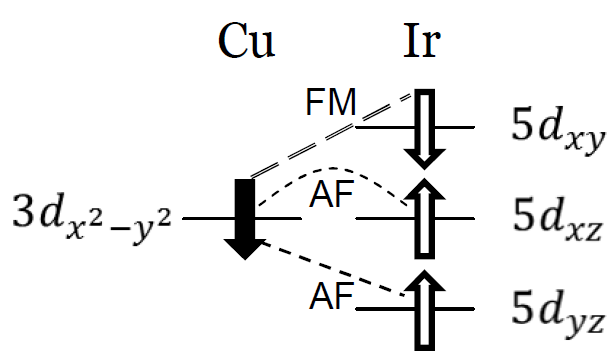}
\caption{\label{fig:exchange}Schematic of unusual exchange pathways between a Cu down spin and an Ir down isospin. All the arrows indicate the real-spin orientations, and open arrows mean partial occupation in the isospin states [Eq.~(\ref{jdis})]. Dashed lines indicate antiferromagnetic (AF) or ferromagnetic (FM) exchange; they cooperate to generate the ferromagnetic spin-isospin interaction. Ir $5d_{xz}$  and $5d_{yz}$ orbitals contribute only to the diagonal part of the exchange processes (see the text), leading to the easy-$z$-axis anisotropy.}
\end{figure}

The different types of magnetic interaction $J_\mathrm{AF}$ and $J_\mathrm{F}$ usually compete with each other. However, we show below from symmetry consideration that this understanding is qualitatively changed by the presence of strong $\lambda$. The three Ir $t_{2g}$ orbitals form a reduced space with an effective orbital angular momentum $l_\mathrm{eff}=1$ with its $z$ component $l_z=0$ for $d_{xy}$ and $l_z=\pm 1$ for $i d_{yz} \pm d_{zx}$ \cite{l1}. The leading energy scales $\lambda=0.44$ eV and $\Delta=0.31$ eV split them into three doublets with the lowest-energy one clearly separated from the other two by $0.58$ and $0.81$ eV, respectively \cite{Sr3CuIrO6:d-d:Liu}. The wave functions of the lowest-energy doublet are of the form \cite{Sr3CuIrO6:supplement}
\begin{equation}
\label{jdis}
|\phi_{0,\eta}\rangle = \frac{1}{\sqrt{2+p^2}}(pd_{xy,\eta}+i d_{yz,\bar{\eta}}+\eta d_{zx,\bar{\eta}}), \;\;\; \eta=\pm,
\end{equation}
where $p$=0.65 ($p$=0,1 for $\Delta=\infty,0$, respectively). With the $z$ component of total angular momentum $j_z=\pm 1/2$, they form an effective $S=1/2$ ``isospin'' on the Ir site  $\vec{s}_{\mathbf{n}}=\sum_{\eta\eta'}{|\phi_{0,\eta}\rangle \vec{\sigma}_{\eta\eta'}^{} \langle\phi_{0,\eta'}|}/2$. Here we have adopted the same coordinate system for all the real spins and isospins in the lattice.
Equation~(\ref{jdis}) reveals that the Ir $5d_{xy}$ and $5d_{xz}$/$5d_{yz}$ orbitals have opposite real-spin orientations ($s_z=\pm 1/2$) in the isospin basis because $s_z=j_z-l_z$.  Therefore, the tendency of the spin in the Ir $5d_{xy}$ ($5d_{xz}$/$5d_{yz}$) orbital to be parallel (antiparallel) to neighboring Cu spins can be satisfied simultaneously via this spin-orbit-coupled Kramers doublet, as illustrated in Fig.~\ref{fig:exchange}. Hence, the $J_\mathrm{AF}$ and $J_\mathrm{F}$ terms counterintuitively cooperate to generate the ferromagnetic interaction between Cu spins and Ir isospins.


Furthermore, we find that by symmetry the Ir $5d_{xy}$ and $5d_{xz}/5d_{yz}$ orbitals make completely different contributions to the isospin-flipping processes [from $\eta=+$ to $-$ or vice versa in Eq.~(\ref{jdis})], in which $d_{xy,+} \leftrightarrow d_{xy,-} $ does not change $l_z$ and $i d_{yz,-} + d_{zx,-} \leftrightarrow i d_{yz,+} - d_{zx,+}$ changes $l_z$ by $\pm 2$. Since $\vec{S}_{\mathbf{n},xy}$ and $\vec{S}_{\mathbf{n},yz}+\vec{S}_{\mathbf{n},zx}$ in Eq.~(\ref{model:t2g}) are $l_z$-preserving operators, the former contributes to the isospin-flipping processes and the latter does not.
This means that the contributions of the $J_\mathrm{F}$ and $J_\mathrm{AF}$ terms in Eq.~(\ref{model:t2g}) to the spin-isospin dynamics are isotropic and anisotropic, respectively. Thus, the strength of the resulting Ising-like exchange anisotropy is positively correlated with the relative importance of the Ir $5d_{xz}/5d_{yz}$ orbitals, which can be substantially enhanced by octahedral titling for increasing $J_\mathrm{AF}/J_\mathrm{F}$ or by octahedral distortion ($\Delta>0$) for increasing $1/p^2$, the relative weight of the Ir $5d_{xz}/5d_{yz}$ orbitals in $|\phi_{0,\eta}\rangle$.

It is noteworthy that in the context of the Kitaev model, an Ising-like ferromagnetic exchange between two Ir isospins due to the simultaneous presence of SOC and Hund's rule coupling was suggested for (Li,Na)$_2$IrO$_3$ \cite{Ir:Jackeli}. We emphasize that the present symmetry-based mechanism for creating the Ising-like ferromagnetic exchange between a Cu spin and an Ir isospin is entirely different because it is independent of Hund's rule coupling.

To describe this strongly anisotropic spin dynamics, we derive from Eq.~(\ref{model:t2g}) the following effective $S=1/2$ spin-isospin Hamiltonian \cite{Sr3CuIrO6:supplement}:
\begin{eqnarray}
\label{final}
H_\mathrm{eff}&=&H^{(0)}+H^{(2)}, \nonumber \\
H^{(0)}&=&-J_1\sum_{\langle \mathbf{m},\mathbf{n}\rangle} {\Big\{ S^x_{\mathbf{m}}s^x_{\mathbf{n}} + S^y_{\mathbf{m}}s^y_{\mathbf{n}} + \gamma_1 S_\mathbf{m}^z s_\mathbf{n}^z \Big\}}, \\
H^{(2)}&=&-J_2 \sum_{\langle\langle\mathbf{m},\mathbf{m}'\rangle\rangle}{\Big\{ S^x_{\mathbf{m}}S^x_{\mathbf{m}'} + S^y_{\mathbf{m}}S^y_{\mathbf{m}'} + \gamma_2 S^z_{\mathbf{m}}S^z_{\mathbf{m}'} \Big\} }, \nonumber
\end{eqnarray}
where $\langle \mathbf{m},\mathbf{n}\rangle$ and $\langle\langle\mathbf{m},\mathbf{m}'\rangle\rangle$ mean the nearest Cu-Ir and Cu-Cu neighbors, respectively. $\vec{S}_{\mathbf{m}}$ is a shorthand notation of $\vec{S}_{\mathbf{m},x^2-y^2}$. $H^{(0)}$ and $H^{(2)}$ contain zeroth-order and second-order terms in perturbation theory, respectively. We find $J_2/J_1 \sim 0.1$; thus, $H^{(0)}$ governs the main physics. $H^{(2)}$ will be shown to have dramatic impact on the atom-specific magnon spectral weight. The strength of the derived exchange anisotropy $\gamma_1-1=(2/p^2)J_\mathrm{AF}/J_\mathrm{F}$ is proportional to $J_\mathrm{AF}/J_\mathrm{F}$ and $1/p^2$, indeed.

\begin{figure}[t]
\includegraphics[width=\columnwidth,clip=true,angle=0]{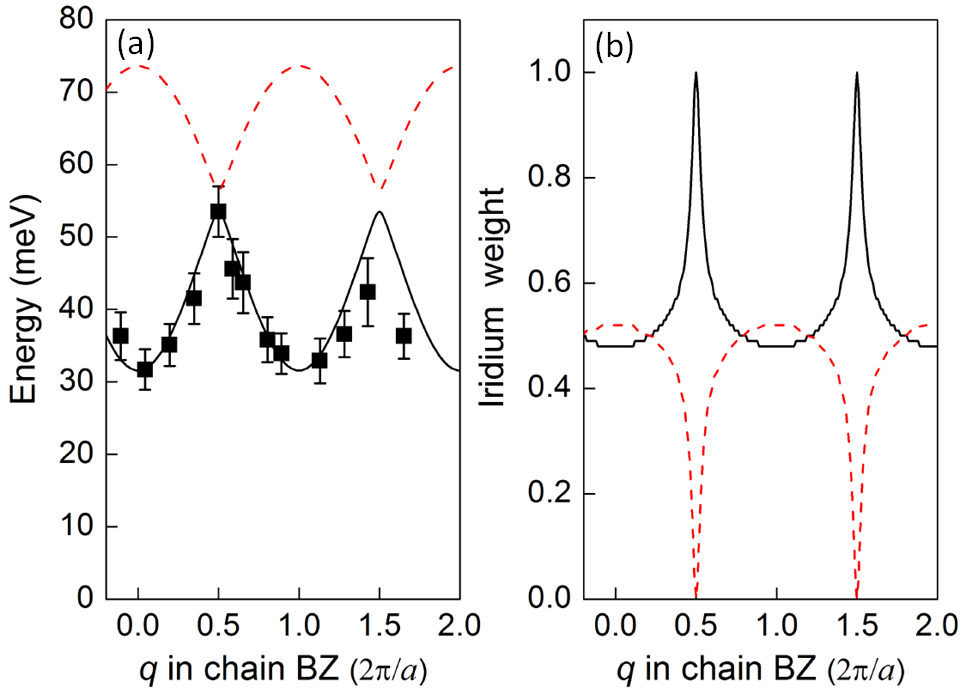}
\caption{\label{fig:spw}(a) Calculated magnon dispersion (lines), compared with the experimental data (solid squares). (b) The Ir partial spectral weights for the lower (solid line) and upper (dashed line) branches in (a). $J_1=21$ meV, $\gamma_1 J_1 = 53.5 $ meV, $J_2=2.4$ meV, and $\gamma_2 J_2=0.6$ meV are used, which satisfy the theoretic constraints \cite{Sr3CuIrO6:supplement}.}
\end{figure}

Note that due to the difference of the (Land\'{e}) $g$ factors for spins and isospins an applied uniform magnetic field along the $z$ axis is seen by the system as a staggered one. Indeed, the isospin's magnetic moment is $\mu_\mathrm{B}(2\mathbf{S}-\mathbf{l})$ \cite{l1} and from Eq.~(\ref{jdis}) it follows that its $g$ factor is $g_J$=$2(p^2-4)/(p^2+2)$=$-2.96$  for the $z$ component, in contrast to $g_S$=2 of the Cu spin. As a result, the saturation magnetization is $\mu_\mathrm{B}|g_J+g_S|/2\simeq0.5 \mu_\mathrm{B}$ per unit cell, close to the experimental value of $0.6\mu_\mathrm{B}$ rather than $2 \mu_\mathrm{B}$ expected for a conventional system containing two unpaired electrons \cite{Sr3(NiCuZn)IrO6:spin-pierls:Nguyen}. Thus our theory provides an explanation of this outstanding puzzle.

$H^{(0)}$ is integrable \cite{spin:Yang,spin:Cloizer}. Its spectrum consists of a fundamental magnon mode and multimagnon bound states. The single-magnon dispersion is
$
\omega(q) = J_1[\gamma_1 - \cos (qa/2)],
$
where $a$ is the nearest Ir-Ir distance and $q$ a momentum along the Cu-Ir chain. This expression coincides with the spin-wave dispersion calculated using the Holstein-Primakoff transformation. Together with $H^{(2)}$, the latter method yields \cite{Sr3CuIrO6:supplement}
\begin{eqnarray}
   \omega_\mp(q)&=& \frac{1}{2} [2 \gamma_1 J_1 + \gamma_2 J_2 - J_2 \cos(qa)] \nonumber\\
   &\mp& \frac{1}{2} \sqrt{[\gamma_2 J_2 - J_2 \cos(qa)]^2 + 4 J^2_1 \cos^2(qa/2)}. \nonumber
\end{eqnarray}
The second-neighbor interaction opens another gap of size $(1+\gamma_2)J_2$ at $q=\pi/a$ in the middle of the band. The lower branch  $\omega_-(q)$ can be fit in good agreement to the experimental data, as shown in Fig.~\ref{fig:spw}(a).

The fact that the upper branch $\omega_+(q)$ spanning from $55$ to $75$ meV is missing in our RIXS data has two implications. One, there exists a channel for the single-magnon excitation at $\omega_+(q)$ to decay into the two-magnon continuum, which starts at about $60$ meV (twice the single-magnon spectral gap), or the multimagnon bound states, which lie around $50$ meV \cite{Sr3CuIrO6:supplement}. Because of strong SOC at Ir sites, lattice irregularities could act as an effective magnetic field applying on the isospins; such a random magnetic field can lead to the decay. This scenario agrees with the previous observation of defect-induced spin-glass behavior in this material \cite{Sr3(CuZn)IrO6:Niazi}.

The other implication concerns the possible insufficiency of the decay near the zone boundary $q=\pi/a$, where $\omega_+(q)$ is very close to $\omega_-(q)$. It is likely that since the RIXS data were taken at the Ir $L_3$ edge, only the Ir weight is visible. The weights of Ir character \cite{Sr3CuIrO6:supplement} in the lower ($-$) and upper ($+$) branch of the spin-wave dispersion $I_\mp(q)$ are
shown in Fig.~\ref{fig:spw}(b). Overall, $I_-(q)$ (solid line) is much larger than $I_+(q)$ (dashed line) and it follows the chain BZ, in agreement with the RIXS data. Note that it is the tiny $J_2$ that dramatically changes the weight distribution; otherwise, $I_\mp(q)\equiv 1/2$ for $J_2=0$. Although this argument alone is not adequate near the zone center, where both $I_\pm(q=0)\simeq 1/2$, the fact that $\omega_+(q=0)$ is the band top implies a strong decay there---both implications are at work to yield the invisibility of $\omega_+(q)$ in our experiment. For a direct experimental verification of the above arguments, we note that $\omega_+(q$=$\pi/a)$ has the full Cu weight [the Cu partial spectral weight is $1-I_\mp(q)$ by the sum rule]. Hence, $\omega_+(q)$ should be detectable near the zone boundary by Cu $L_3$ edge RIXS \cite{RIXS:La2CuO4:Dean}. This will be pursued in the future \cite{note:RIXS}.

In summary, we have presented a combined experimental and theoretical study of the unusual ferromagnetism in Sr$_3$CuIrO$_6$. We have revealed strongly anisotropic spin dynamics in this material and found that an unusual exchange anisotropy generating mechanism, namely, strong ferromagnetic anisotropy arising from antiferromagnetic superexchange, is generally present in the systems with edge-sharing Cu$^{2+}$O$_4$ plaquettes and Ir$^{4+}$O$_6$ octahedra. Our results demonstrate that mixed $3d-5d$ compounds can generate distinct exchange pathways and thus novel magnetic behaviors that are absent in pure $3d$ or $5d$ compounds.

The work at Brookhaven National Laboratory was supported by the U.S. Department of Energy (DOE), Division of Materials Science, under Contract No. DE-AC02-98CH10886. Use of the Advanced Photon Source was supported by DOE, Office of Science, Office of Basic Energy Sciences, under Contract No. DE-AC02-06CH11357. T.F.Q. and G.C. were supported by the NSF through Grant No. DMR-0856234.


\begin{thebibliography}{28}
\bibitem [{not()}]{note:RIXS}%
  \BibitemOpen
  \href@noop {} {}\bibinfo {note} {To date, the energy resolution of Cu $L_3$
  edge RIXS is about 130 meV for practical use, which is not good enough to
  discern the magnetic excitations $\sim 60$ meV in this material. Facilities
  expected to operate with 10 meV resolution are under construction, for
  example, at Brookhaven National Laboratory as part of NSLS II.}\BibitemShut
  {Stop}%
\end{thebibliography}
%

\end{document}